\documentclass[12pt]{iopart}

\usepackage{latexsym,graphicx,amssymb,fancyhdr}

\newcommand{\ket}[1]{| #1 \rangle}
\newcommand{\bra}[1]{\langle #1 |}
\newcommand{\braket}[1]{\langle #1 \rangle}

\newcommand{\pl}{photoluminescence}
\newcommand{\Pl}{Photoluminescence}

\begin{document}

\title[Characterization of dynamical regimes and entanglement sudden death...]{Characterization of dynamical regimes 
    and entanglement sudden death in a microcavity quantum 
    - dot system}

\author{Carlos A. Vera, Nicol\'as Quesada M.$^*$, Herbert Vinck-Posada, Boris A. Rodr\'iguez}

\address{Universidad de Antioquia, Instituto de F\'isica,
    Medell\'in, AA 1226 Medell\'in, Colombia}
\ead{$^*$nquesada@pegasus.udea.edu.co}
\begin{abstract}
The relation between the dynamical regimes (weak and strong
  coupling) and entanglement for a dissipative quantum - dot
  microcavity system is studied. In the framework of a phenomenological temperature model an analysis in both, temporal
  (population dynamics) and frequency domain (\pl) is carried out
  in order to identify the associated dynamical behavior. The Wigner
  function and concurrence are employed to quantify the entanglement
  in each regime. We find that sudden death of entanglement is a
  typical characteristic of the strong coupling regime.

\end{abstract}

\submitto{\JPCM}
\maketitle

\section{\label{sec01}Introduction}
In the last few years, the study of microsystems that involve the
interaction between an active medium and confined light, has made
possible the observation of interesting phenomena in two dynamical
regimes: weak and strong coupling \cite{Laussy-book, Yamamoto-book,
  Gayral-th}.  In the first regime, spontaneous emission control
was successfully realized (Purcell effect) \cite{Santori, whittaker};
in the second one, several groups are now searching for coherent
polaritonic phenomena such as lasing \cite{bloch} or condensation
\cite{dang, Deng02}, which can open applications in quantum
information and quantum optics \cite{nature,Deng06, Yamamoto07, Forchel}.  In
addition, some recent theoretical works have shown that the dynamical
properties for a coupled quantum dot - cavity system, may be described
using a simple dissipative model \cite{Tejedor04, Vinck}.\\

Our purpose in this work is to study the relations between weak and
strong coupling regimes with the dynamical exciton - photon
entanglement, by using the concurrence measure and the Wigner
quasiprobability function. 
Despite the fact that the Wigner function
depends only on the photonic state, it has been shown that it can be
used as a qualitative criterion for determining the separability of
the quantum exciton - photon state \cite{Vogel}.
In reference \cite{teje} a similar study has been carried
out but studying 2 quantum dots and only in the stationary limit.
\\

The paper has been written as follows: section \ref{sec02} contains
the theoretical framework supporting our model, in section
\ref{sec02:sub01} we describe the system and the dissipative processes
that models its dynamics using a master equation. In section
\ref{sec02:sub02} we explain how the \pl~ spectrum is obtained by
using the quantum regression theorem. In section \ref{sec02:sub03},
using a simple phenomenological model we include temperature
effects in the quantum dot gap. In section \ref{sec02:sub04} we review
the concurrence and Wigner function concepts, and their connection
with the photon - exciton entanglement. Hence, the dynamical regimes
are characterized employing the numerical integration results of the
master equation for two different cutoff conditions in the photon
number, in both cases, showing a good agreement with experimental
data. Once the regime is characterized, a dynamical description of the
entanglement is analyzed and we show collapses and revivals of this
quantity as a function of the dissipative parameters involved in the
model. Furthermore, we establish the usefulness of the Wigner function
criterion to detect separability in a multi-state system where the
concurrence criterion can not be used. Finally, some conclusions are given in
the last section.\\

\section{\label{sec02}Theoretical background}

We are interested in the evolution of a quantum dot interacting with a
confined mode of the electromagnetic field inside a semiconductor microcavity. In
these systems, quantum states associated to the matter excitations,
the so-called excitons, are bound states resultant of the Coulomb interaction
between electrons in the conduction band and holes in the valence band. This
quasi-particle exhibits a complete discrete excitation spectrum,
however, in this work we will only consider the first two levels of
this set, the ground $\ket{G}$ (no excitation, i.e, electron in the
valence band) and excited $\ket{X}$ (exciton) states. This assumption
is based upon the fact that the ground and first excited states of a
multilevel model involving Coulomb interaction, are mainly filled in
the dynamical evolution of the system \cite{gonzalez}. A possible
experimental realization of this model, can be implemented in
a pumped system with polarized light \cite{stace} and slow spin flip
mechanisms. The photonic component will be
treated as a single electromagnetic mode. The
validity of this usual assumption is subjected to the existence of
well separated modes in energy inside the cavity
\cite{gerard}. The last condition amounts to say that for example the radii of
the micropillar is small since the energy separation of the modes increases when the radii is decreased.
We will employ a Fock state basis $\ket{n}$,
which ought to be truncated in order to implement computationally the
dynamics.\\
Light-matter interaction is described by the  Jaynes-Cummings Hamiltonian,

\begin{equation}
  H =  \omega |X \rangle \langle X| +  \omega_0 a^\dag a +
   g(\sigma a^\dag + a \sigma^\dag),
\end{equation}

where $\sigma = |G\rangle \langle X|$ and $\sigma^\dag = |X\rangle
\langle G|$ are exciton ladder operators and $a$ ($a^\dag$) is the
annihilation (creation) operator for photons. $\omega$ and
$\omega_0$ are the exciton and photon energy, respectively, and
$g$ is the coupling constant and we have set $\hbar=1$.\\ We also define the detuning between the
exciton and photon frequency as $\Delta = \omega -\omega_0$. Under
these considerations the system is Hamiltonian and completely
integrable \cite{Scully}. However, real physical systems are far away
from this simple description; dissipative effects play an important
role in the evolution of the system. Indeed, if no losses were
considered in the system, no measurements could be done since light
would remain always inside the microcavity. 

\subsection{\label{sec02:sub01}Dynamics}

The whole system-reservoir hamiltonian can be splitted 
in three parts. One is for the system we are considering, namely, the photons of the cavity 
and the exciton. The second one is the hamiltonian of the reservoirs, which is made of 
electron-hole pairs, photons and phonons and finally a third part which is a bilinear coupling
between the system and the reservoirs. The explicit form of the system-reservoir interaction for this model can be found in 
\cite{Tejedor04}.
After tracing over the external reservoirs degrees of freedom and assuming the validity of the
Born-Markov approximation, which requires weak coupling between the system (exciton+cavity photons) and the reservoirs one arrives to 
a master equation. It has recently been found \cite{quiroga} that non-markovian dynamics is relevant for high pumping intensities. \\
The master equation we have found accounts for three different processes,
namely, coherent emission ($\kappa$), external pumping ($P$) and
spontaneous emission ($\gamma$). The  master equation we shall consider is \cite{Tejedor04,walls,Gerry}:

\newpage

\begin{eqnarray}\label{mastereq}
  \frac{d\rho}{dt} &=& i[\rho,H] + \frac{\kappa}{2}
  (2a\rho a^\dag - a^\dag a\rho -\rho a^\dag a) +
  \frac{\gamma}{2}(2\sigma \rho \sigma^\dag - \sigma^\dag \sigma \rho
  - \rho \sigma^\dag \sigma) \nonumber \\ && +
  \frac{P}{2}(2\sigma^\dag \rho \sigma - \rho \sigma \sigma ^\dag - \sigma \sigma^\dag \rho
  ).
\end{eqnarray}

\begin{figure}[!ht]
  \begin{center}
    \includegraphics{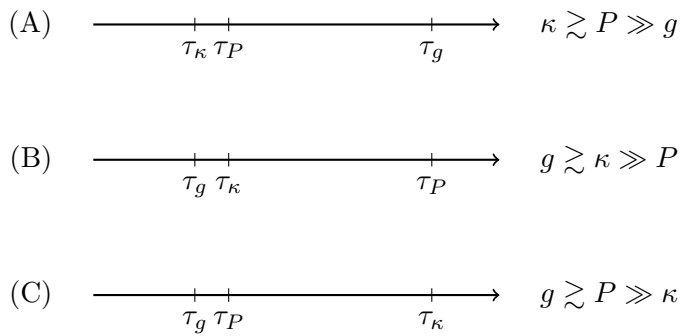}
  \end{center}
  \caption{\label{timescales} Schematic diagram of some relations
    among the typical time scales associated with the effective rates
    involved in the master equation. Cases (A) and (B) show situations
    in which the basis containing up to one photon is enough for
    describing the dynamics. On the other hand, case (C) shows a set
    of parameters where this description fails, and the requirement
    for a larger basis arises. See the text for additional details.}
\end{figure}

By changing the values of the free parameters in this model, two
different dynamical regimes are reached: weak and strong
coupling. Possible transitions between these two regimes, can be
achieved as loss and pump rates are modified. The dynamical behavior
of the system as well as the size of the basis employed  are 
governed by the competition of the time scales involved in the
model. The time scales associated with the non conservative processes included  are
given by: $\tau_P = 1/P$, $\tau_{\kappa} = 1/\kappa$ and
$\tau_{\gamma} = 1/\gamma$, whereas, the interaction time scale is
$\tau_g = 1/g$. We address now to the interpretation of these time
scales. In figure \ref{timescales} three possible cases are shown whose
dynamics can be described using a basis including up to one photon
(cases (A) and (B)) and more than one photon (case (C)).

In case (A), for instance, the relation $\kappa \gtrsim P \gg g$ holds,
and the system is operating in weak coupling regime. When an
\emph{exciton} is pumped during the typical time scale $\tau_P$, the
elapsed time until it recombines into a \emph{photon} is given by the
(largest) scale $\tau_g$. Bear in mind that throughout this period no
further excitation can be done over the quantum dot, because the
Pauli exclusion principle does not allow an additional
excitation. The photon generated in this way, quickly leaves the
cavity due to the small time scale $\tau_\kappa$ in which it can inhabit the cavity.
The latter mechanism applies for any photon living the cavity.
 
Hence we see that there is
no chance for Rabi oscillations in the dynamical evolution. If the
initial condition for the electromagnetic field is the vacuum state,
the photon mean number is never expected to be greater than
one.\\ A similar reasoning can be done in the case (B), where
$g\gtrsim \kappa \gg P$. In this case each photon created by the
interaction can be re-absorbed and produce Rabi oscillations (i.e.,
strong coupling), before it leaves the cavity. A
basis including up to one photon is enough to describe the system because the characteristic
time related to losses is by much smaller than the pump one. The
important issue here is that the dynamical situation is clearly
different to the previous one.\\Finally, in case (C), where $g \gtrsim
P \gg \kappa$, we have a situation in which the photons can be
efficiently stored within the cavity. Indeed, since the coherent
emission rate is small, its characteristic time scale is large enough
compared with the associated excitonic pump and interaction
rates. Photons created  by exciton recombination remain in the
cavity a long time before they leak through the cavity mirrors. This
case is a clear example where the multiphoton basis must be
implemented.\\

\subsection{\label{sec02:sub02}\Pl}

The Fourier transform of the first order correlation function directly
gives the \pl~ spectrum of the system \cite{Scully},

\begin{equation}\label{spectequation}
  S(\omega,t) \;\propto\; \int_{-\infty}^{\infty} \langle
  a^\dag(t+\tau) a(t) \rangle \e^{i\omega \tau}d\tau.
\end{equation}

\noindent Note that in this expression a knowledge of the
time-dependent expected value for the product of creation and
annihilation operators, is needed in order to compute the \pl~
spectrum. However, non analytical expression for such expectation
value is available for our system. This problem can be solved by
representing the operators $a$ ($a^\dag$) in the interaction picture,
and then using the quantum regression theorem \cite{walls}. This
theorem states that given a set of operators $O_J$ satisfying,

  \begin{equation}
    \frac{d}{d\tau} \braket{O_j(t+\tau)} = \sum_k L_{jk} \braket{O_k(t+\tau)},
  \end{equation}
  
  then
  
  \begin{equation}
    \frac{d}{d\tau} \braket{O_j(t+\tau)O(t)} = \sum_k L_{jk}
    \braket{O_k(t+\tau)O(t)}.
\end{equation}

for any operator $O$. The validity of this theorem holds
whenever a closed set of operators are involved in the dynamics in the
Markovian approximation. Unfortunately, representing creation and
annihilation operators in the interaction picture does not lead to a
complete set. It is necessary to add two new operators in order
to close the system. The final set of equations for the operators are:

\begin{eqnarray}
  &&a^\dag_{Gn}(t) = \ket{Gn+1}\bra{Gn}\e^{i(\omega-\Delta) t} \nonumber \\
  &&a^\dag_{Xn}(t) = \ket{Xn+1}\bra{Xn}\e^{i(\omega-\Delta) t} \nonumber \\
  &&\sigma^\dag_{n}(t) = \ket{Xn}\bra{Gn}\e^{i\omega t} \nonumber \\
  &&\zeta _n(t) = \ket{Gn+1}\bra{Xn-1}\e^{i(\omega-2\Delta) t}.
\end{eqnarray}
In the framework of the previous discussion regarding the dynamical
time scales, we can consider the simplest model involving all
dissipative and interaction processes, that is, we shall study a
problem having just three levels: $\ket{G0}$, $\ket{X0}$ and
$\ket{G1}$. Taking into account this consideration we only need to
define two operators whose dynamical equations are:

\begin{eqnarray}
  &&\frac{d}{dt}\braket{a_{G0}^\dag(t)} =
  -\left(\frac{\kappa}{2}+P-i(\omega-\Delta)
  \right) \braket{a_{G0}^\dag(t)} +ig \braket{\sigma_0^\dag(t)}
  \nonumber \\
  && \frac{d}{dt}\braket{\sigma_0^\dag(t)} = ig\braket{a_{G0}^\dag(t)} -
  \left(\frac{P+\gamma}{2}-i\omega \right) \braket{\sigma_0^\dag(t)},
\end{eqnarray}

\noindent thus by using the quantum regression theorem we can write
the dynamics for the delayed operators as,
\begin{eqnarray}
  \dot{X} &=& \frac{d}{d\tau}
  \left(\begin{array}{c}
      {\braket{a_{G0}^\dag(t+\tau)}} \\
      {\braket{\sigma_{0}^\dag(t+\tau)}}
    \end{array}\right) =
  \nonumber \\ &&
  \left( \begin{array}{cc}
      -(\frac{\kappa}{2}+P) +i(\omega-\Delta) & ig \\
      ig & -\frac{P+\gamma}{2}+i\omega
    \end{array}\right)
  \nonumber   \left(\begin{array}{c}
      \braket{a_{G0}^\dag(t+\tau)} \\
      \braket{\sigma_{0}^\dag(t+\tau)}
    \end{array}\right) = \mathcal{A}X.
\end{eqnarray}

\noindent The last linear system has the following formal solution

\begin{equation}
  X(t+\tau) = \e^{\mathcal{A}\tau}X(t) = \mathcal{B}\e^{\Lambda \tau}
  \mathcal{B}^{-1} X(t),
\end{equation}

\noindent where $\mathcal{A}=\mathcal{B}\Lambda \mathcal{B}^{-1}$ and
$\Lambda=\mathrm{diag}\{\lambda_+,\lambda_- \}$ is a diagonal matrix
containing the eigenvalues of $\mathcal{A}$.  %Since the emission

The eigenvalues $\lambda_\pm$ are directly related with
the peaks $\omega_\pm$ and widths $\Gamma_\pm$ of the spectrum,

\begin{equation}\label{lambda}
  \omega_\pm + i \Gamma_\pm = i \lambda_\pm.
\end{equation}

\subsection{\label{sec02:sub03}Temperature effects}

In order to make a more realistic description of the experimental data
we also include a temperature dependence in the model. It is important
to explicitly point out that the developed master equation describes
the system at zero temperature, that is, the states of the reservoirs considered
in the derivation of the master equation (\ref{mastereq}) are zero temperature states.
This assumption can be justified as follows.

First notice that when a finite temperature reservoir of photons 
is considered in the derivation of the master equation one obtains two Lindblad  terms \cite{petruccione}.
One that accounts for thermally induced absorption and that is proportional to the average number of photons of the reservoir and 
another term that is proportional to the average number of photons of the reservoir \emph{plus one} and that accounts for
spontaneous emission. 

The average number of photons of a thermal reservoir at temperature $T$ is $1/(e^{ \omega_0 / k_B T}-1)$.
The standard experimental values for the photon energy in a micropillar are $ \omega_0 \sim 1$ meV whereas for temperatures of the order of $10^2$ K the thermal energy $k_B T \sim 1$ meV and thus the average number of photons is $N(\omega) \sim  e^{-10^3} \sim 10^{-400}$. We see that effects due to finite temperature  in the master equation for the system considered here are quite small.\\

Summarizing, the above discussion, we can add thermal
effects on the system through its effective parameters, such as the
quantum dot energy gap and the cavity refractive index, without
considering the fundamental issues introduced into the
model by the dephasing processes. First, let us consider the
temperature photon energy dependence. Usual microcavities are built 
from GaAs/AlGaAs layers, for these cavities the resonant wavelength
depends on the refractive index as $\lambda =
\lambda_{\mathrm{air}}/n$, where $ \lambda_{\mathrm{air}}$ is the
light wavelength in vacuum. We can then modify the resonant frequency
by changing the refractive index $n(T)$ \cite{Forchel}. Experiments have
shown that this index has an almost linear temperature dependence
within the range from 0 K up to few hundred Kelvin.
%, thus, temperature variations allow us to change the cavity energies.
 In reference \cite{Blakemore}
has been shown that the refractive index can be modeled with the simple formulae:

\begin{equation}\label{ch02:sec01:eq04pp}
 n(T) = n_0(1+aT).
\end{equation}
Where $a\sim 10^{-5}$ K$^{-1}$. It is seen that the corrections to the
wavelength due to the temperature are rather small, and because of this will not be considered in this work.
On the other hand, the temperature effects, related to the active
medium can be included through the energy gap in the quantum dot. We
shall use here the Varshni model \cite{vurgaftman01}, which
fits the band gap thermal dependence in the low temperature region. A
more detailed discussion about the validity of this model, can be
found in reference \cite{passler},

\begin{equation}\label{ch02:sec01:eq04p}
   \omega(T) = E_G(0) -\frac{\alpha T^2}{\beta +T},
\end{equation}

In this work we will use two different sets of experimental data for
InGaAs quantum dots, for which the Varshni parameters were fitted.

\subsection{\label{sec02:sub04}Concurrence and Wigner Function}

Determining entanglement between two quantum mechanical systems is a
complicated task when the systems involved have several degrees of
freedom, that is, when the basis representing the density matrix has
more than two pairs of states. However, if each system is completely described
with two levels (i.e. each system is a \emph{qubit}) the situation
becomes easier; indeed, the \emph{entanglement} can be obtained from
the \emph{concurrence} expression \cite{Wooters},

\[ C(\rho) = \max \{0,\lambda_1-\lambda_2-\lambda_3-\lambda_4\},\]

where $\{\lambda_i\}$ are the square root of the decreasingly ordered
eigenvalues of the positive-definite non-Hermitian matrix $\rho
\tilde{\rho}$ with,

\[ \tilde{\rho} = (\sigma_y \otimes
\sigma_y)\rho^* (\sigma_y \otimes \sigma_y), \]

and $\rho^*$ is the complex conjugated matrix representation of $\rho$
in some suitable basis.\\

For instance, note that if our  system is described by using the 
basis $\{ |G\rangle, |X\rangle\}
\otimes \{| 0 \rangle , |1\rangle \}$, (where $|0 \rangle ,|1\rangle $ are photon Fock states), the exciton - photonic field system can be thought as two interacting qubits. Once a
(numerical) solution of the master equation is obtained, it is
straightforward to compute the concurrence following the previous recipe. It is to be noticed that the validity of this basis is
constrained to the dynamical regime where this cutoff holds. When the
basis is larger than the one considered earlier, the concurrence measure is not
applicable and hence a new criterion must be established. Despite
the Wigner function only depends on the photon state, it has been
demonstrated that it yields qualitative information about the
separability between the exciton and photon parts of the global
quantum state \cite{Vogel}. The Wigner function for the photonic field
can be easily computed as \cite{Davidovich},

\begin{equation}
  W(\alpha) = 2\;\mathrm{Tr}_P[D(-\alpha)\rho^{(P)} D(\alpha) P_f],
\end{equation}

\noindent where $P_f=\e^{i\pi a^\dag a}$ is the field parity operator
and $D(\alpha) = \e^{\alpha a^\dag -\alpha^* a}$ is the displacement
operator, $\rho^{(P)}$ is the density operator of the photons and $ \mathrm{Tr}_P$ 
is the trace operation in Fock space.
 Our system involves both excitonic and photonic
states, hence this definition is not directly applicable, and a
previous differentiation among the possible combinations has to be
done, that is, we have to separately consider the Wigner matrix
elements $W_{ij}$ \cite{Vogel},
 
\[ W_{ij}(\alpha) = 2\;\mathrm{Tr_P}[D(-\alpha)\bra{i}\rho\ket{j}
  D(\alpha) P_f],
\] 

\noindent where indexes $i,j$ run over excitonic states
$\{\ket{X},\ket{G}\}$, and $\rho$ is the density operator of the whole system. Notice that $\bra{i}\rho\ket{j}$ is an operator
that acts only in the state space of the photons so that the operations in the above equation are well defined.

Note that if the system is separable at a given time then the matrix elements  $W_{XX}$
and $W_{GG}$ have the same shape in phase space, leading 
to a separable system, which is itself related to a vanishing
concurrence situation. Indeed, recognizing  that each Fock state has a
well defined signature within the phase space, it is possible to
identify the predominant photonic state by observing the Wigner
function. A more detailed analysis shows that if the quantum state is
separable,
\[ \rho =\sum_I \rho_I ^{(P)} \otimes \rho_I ^{(X)},\]
\noindent then the corresponding Wigner function of the system can be
written as \cite{Vogel}

\begin{equation}
  W_{ij}(\alpha) = \rho^{(X)}_{ij} W(\alpha),
\end{equation}

\noindent where $W(\alpha)=\sum_i W_{ii}(\alpha)$. Therefore, if the system is separable at a
given time all the matrix elements $W_{ij}$ of the Wigner matrix  have the
same shape for every $i,j$. In opposition to the
quantitative concurrence measure, the Wigner function only shows qualitative
information about the entanglement.\\

\section{\label{sec03} Results}

\subsection{Weak coupling regime \label{sec03:WC}}

In the weak coupling (WC) regime the relation \cite{Gerry}

\begin{equation}\label{wccond}
  16g^2 < (\kappa -\gamma)^2,
\end{equation}

\noindent holds, hence by selecting a strong enough emission rate
WC is obtained. We choose $ g=15$ $\mu$eV, $\gamma = 1$ $\mu$eV, $\kappa=85$ $\mu$eV
and $ P=20$ $\mu$eV corresponding to the case (A) of figure
\ref{timescales}. As we stated before, this situation can be described
with the reduced basis including up to one photon.\\
We stress that relation (\ref{wccond} )
is derived under the assumption that there is no incoherent pumping 
over the system. The effects of the pumping have been studied in detail in reference \cite{teje1}.

\begin{figure}[!ht]
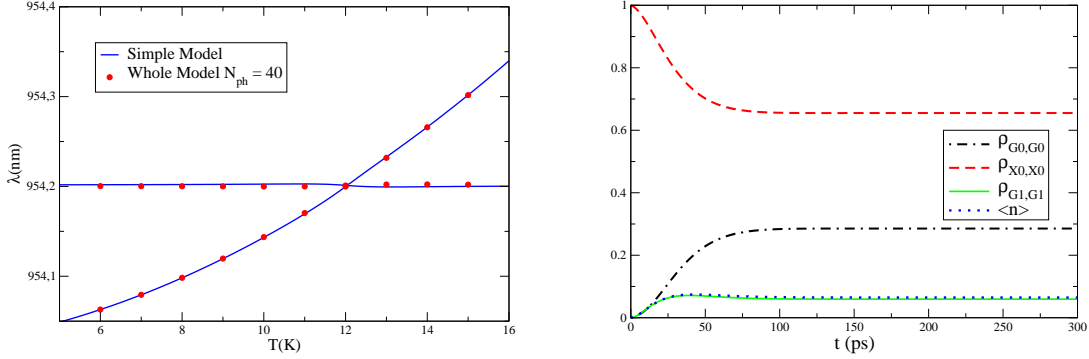

  \begin{minipage}{0.5\textwidth}
    \includegraphics[height=0.60\textwidth,angle=0]{WC.eps}
  \end{minipage}
  \ \hfill
  \begin{minipage}{0.5\textwidth}
    \includegraphics[height=0.60\textwidth,angle=0]{figure1.eps}
  \end{minipage}
  \caption{\label{fig:1} (left panel) Crossing emission peaks in the
  WC regime computed with a simplified model (solid line), the whole
  model using a 40 photonic level basis (squares). (right panel) Time evolution of the populations and the average photon number in the WC
  regime, computed using the simplified model, $ g=15$ $\mu$eV, $\gamma = 1$ $\mu$eV, $\kappa=85$ $\mu$eV
and $ P=20$ $\mu$eV. The average photon number was computed using 40 Fock states.}
\end{figure}

We have taken a photon decay rate of $\kappa=85$ $\mu$eV and a cavity mode of frequency
$\omega=1296.11$ meV, so that the quality factor of the cavity we are considering is $Q \approx 15000$. 
The spectra for this set of parameters is shown in figure \ref{fig:1}.
These values were chosen following the results of reference \cite{Yamamoto07} where strong coupling in a single quantum dot -
microcavity system is reported.
\\
In figure \ref{fig:1} it is seen that at approximately $T=12 K$ the peaks of the
cavity and exciton coincide. 
The parameters used for the temperature dependence where $E_G(0)=1299.6$ meV, $\alpha=0.81$ meV/K, 
$\beta = 457.6$ K.\\
\\

We pointed out that weak coupling regime could be reached by setting an
emission rate higher than the coupling constant. Due to the small time the photons
spend in the cavity the chance of them to interact with the exciton is quite small
and because of this no oscillations appear in the populations as seen in Figure~\ref{fig:1}.

\subsection{Strong coupling}

The strong coupling (SC) regime is even more interesting since it
enables the existence of \emph{polaritons}. From an experimental
point of view it is more complicated to obtain due to the fine scales
of the variables involved. This regime is also studied with the same
three level model described above and with a basis of 40 Fock states
but now taking, $ g=35\;\mu$eV, $ \kappa = 25\;\mu$eV, $ \gamma =
1\;\mu$eV, $ P=1\;\mu$eV, $E_G(0)$= 1299.6 meV, $\alpha$= 0.81 meV/K,
$\beta $= 457.6 K and a cavity resonant frequency of $\nu=1299.35$
meV. \\
It is clear that in this regime oscillations in the populations are observed since 
the time a photon can live in the cavity is long enough for it to interact with the exciton, 
i.e., $1/g=\tau_g >\tau_\kappa=1/\kappa$. \\
\\

\begin{figure}[!ht]
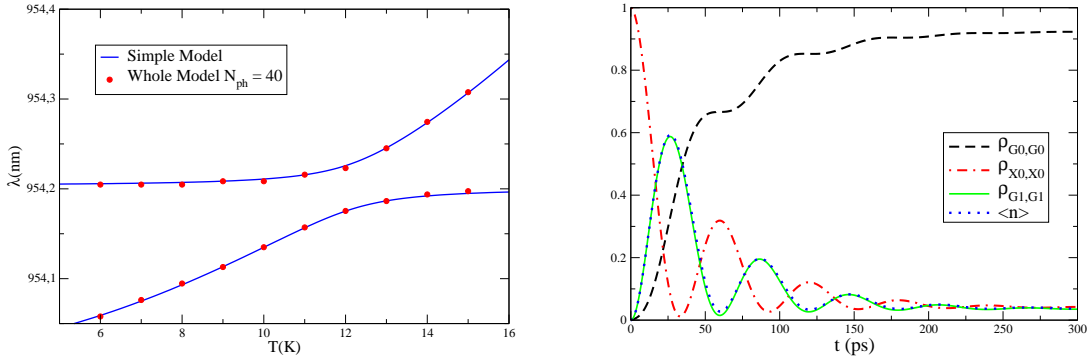

  \begin{minipage}{0.5\textwidth}
    \includegraphics[height=0.6\textwidth,angle=0]{SC.eps}
  \end{minipage}
  \ \hfill
  \begin{minipage}{0.5\textwidth}
    \includegraphics[height=0.6\textwidth,angle=0]{figure2.eps}
  \end{minipage}
  \caption{\label{fig:5} (left panel) Emission peaks in the SC
    regime. Anticrossing can be observed at $T = 12$ K. Solid line
    was computed by using the simple three level model, meanwhile
    dots are the peaks obtained with a basis involving 40 Fock
    states. (right panel) Time evolution of populations and the average photon number for the strong
    coupling regime, again only the average photon number was computed with 40 Fock states. Differences with the weak coupling regime are
    clear.}
\end{figure}

Peaks positions are shown in fig. \ref{fig:5}.

Note that for detunings far
from the resonance, a small shift of the cavity peak appears, this
effect is not observed in the WC regime. This shift is due to the strong interaction between matter and light.\\ The closest approximation of the two emission peaks (cavity and excitonic) is reached when $T=12$
K, where the separation (Rabi splitting) is

\[ R = 68.89\;\mu \mathrm{eV} ,\]

\noindent Another important fact which enables us to conclude that we are
dealing with a system in SC regime is that the following condition holds

\[ R \approx  2g \]

That is the Rabi splitting is approximately two times the coupling constant \cite{Gerry}.\\
Finally to determine if the reduced basis was enough to describe the system we computed 
the average photon number $\braket{n_{40}}$ using a basis of 40 Fock states. This is information is
in the right panels of figures \ref{fig:1} and \ref{fig:5} and can be compared with the average photon
number using the reduced basis: $\braket{n_1}=\rho_{G1G1}$. It is seen that both descriptions 
are in complete agreement.

\begin{figure}%[!ht]

  \begin{minipage}{0.5\textwidth}
    \includegraphics[width=0.95\textwidth, angle=0]{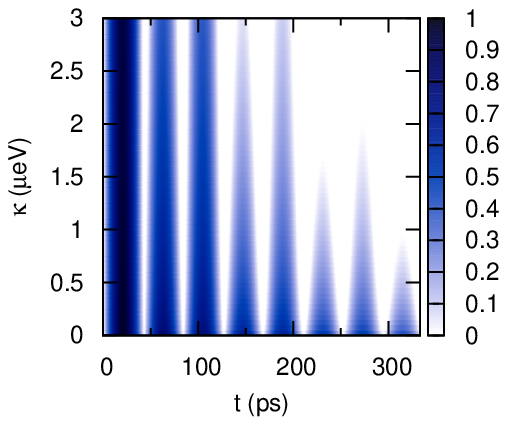}
    \put(-210,160){(a)}
  \end{minipage}
  \ \hfill
  \begin{minipage}{0.5\textwidth}

    \includegraphics[width=0.95\textwidth, angle=0]{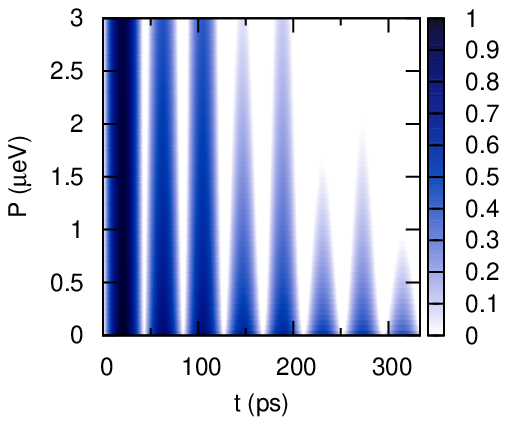}
\put(-210,160){(b)}
  \end{minipage}
  \\ 
\\
\\

  \begin{minipage}{0.5\textwidth}
    \includegraphics[width=0.95\textwidth, angle=0]{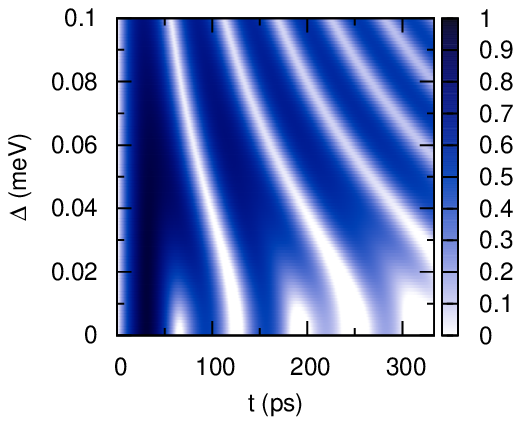}
\put(-210,160){(c)}
  \end{minipage}
  \ \hfill
  \begin{minipage}{0.5\textwidth}
    \includegraphics[width=0.95\textwidth, angle=0]{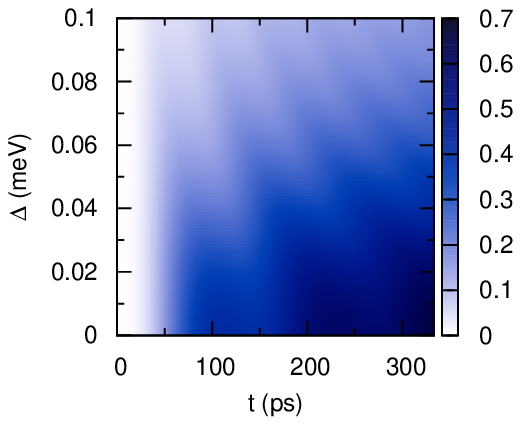}
\put(-210,160){(d)}
  \end{minipage}
  \\
\\
\\

  \begin{minipage}{0.5\textwidth}
    \includegraphics[width=0.85\textwidth, angle=0]{grafica1.eps}
\put(-220,130){(e)}
  \end{minipage}
  \ \hfill
%fig12
  \begin{minipage}{0.5\textwidth}
    \includegraphics[width=0.85\textwidth, angle=0]{grafica2.eps}
\put(-220,130){(f)}
  \end{minipage}\caption{\label{fig:7} 
Concurrence time evolution as a function
  of the dissipative parameters: (a) changing $\kappa$ and fixing
  $ P=3\;\mu$eV, $\Delta=0$ , (b) changing $P$ and fixing
  $ \kappa =3\;\mu$eV, $ \Delta = 0$  and (c) changing $ \Delta$
  and keeping $ P=2\;\mu$eV, $ \kappa=3\;\mu$eV . In (d) time
  evolution of linear entropy ($M(\rho)=1-\Tr(\rho^2)$) for the last set of parameters. In figures (e) and (f) we plot the concurrence ($C(\rho)$), linear entropy ($M(\rho)$),  $\sqrt{\rho_{G0G0} \rho_{X1X1}}$ and $|\rho_{X0G1}|$ for  $\Delta = 0$ eV and $ \Delta = 0.1$ eV respectively with $ \kappa=3\;\mu$eV, $ P=2\;\mu$eV. Dynamics was solved with the initial condition $\rho_{X0X0}=1$ and the coupling constant $ g=25\;\mu$eV in all cases.}

\end{figure}

\subsection{Concurrence}

\begin{figure}[ht]
  \begin{minipage}{0.5\textwidth}    
    \includegraphics[width=0.85\textwidth,angle=-90]{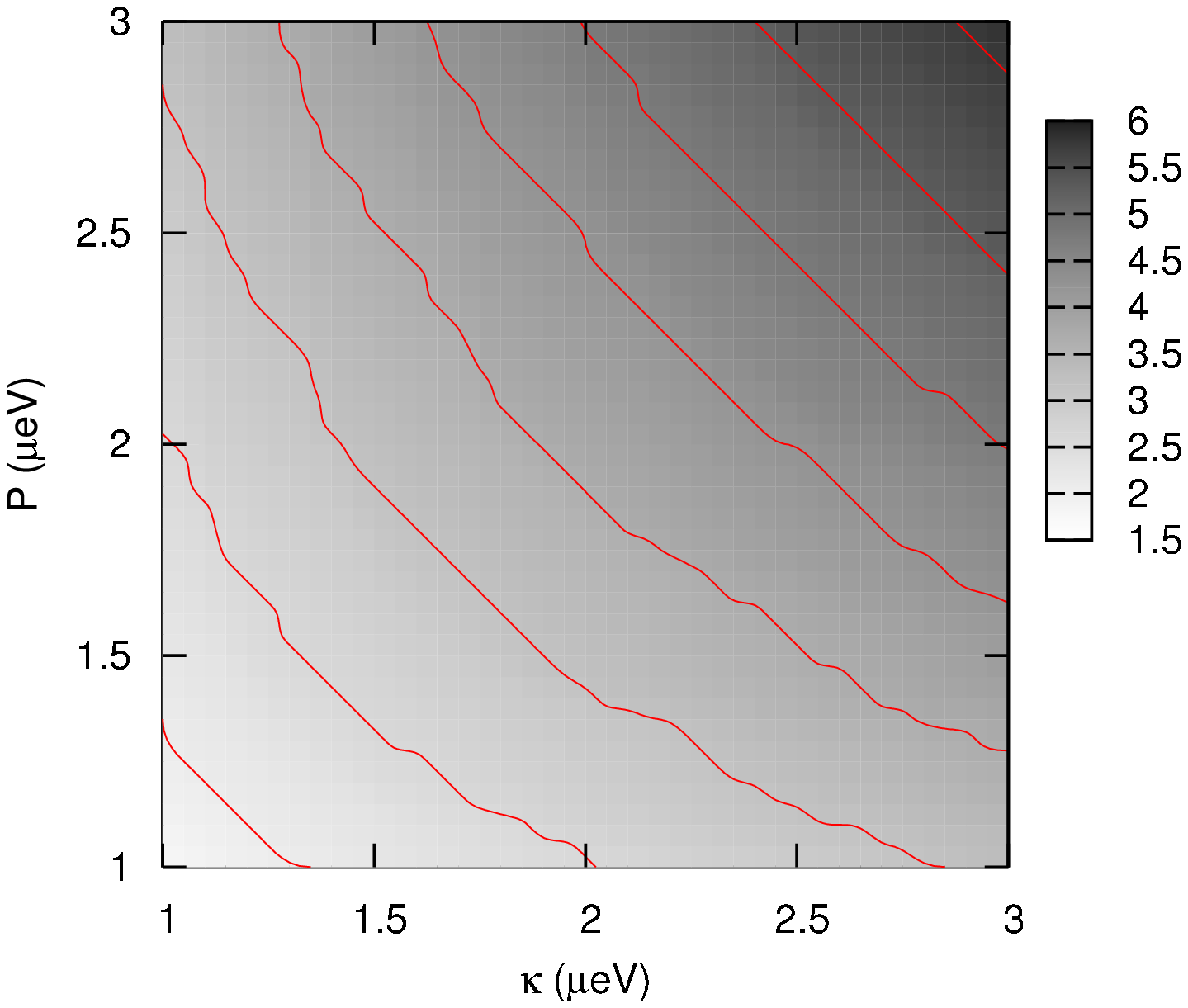}
  \end{minipage}
  \ \hfill
  \begin{minipage}{0.5\textwidth}    
    \includegraphics[width=0.85\textwidth,angle=-90]{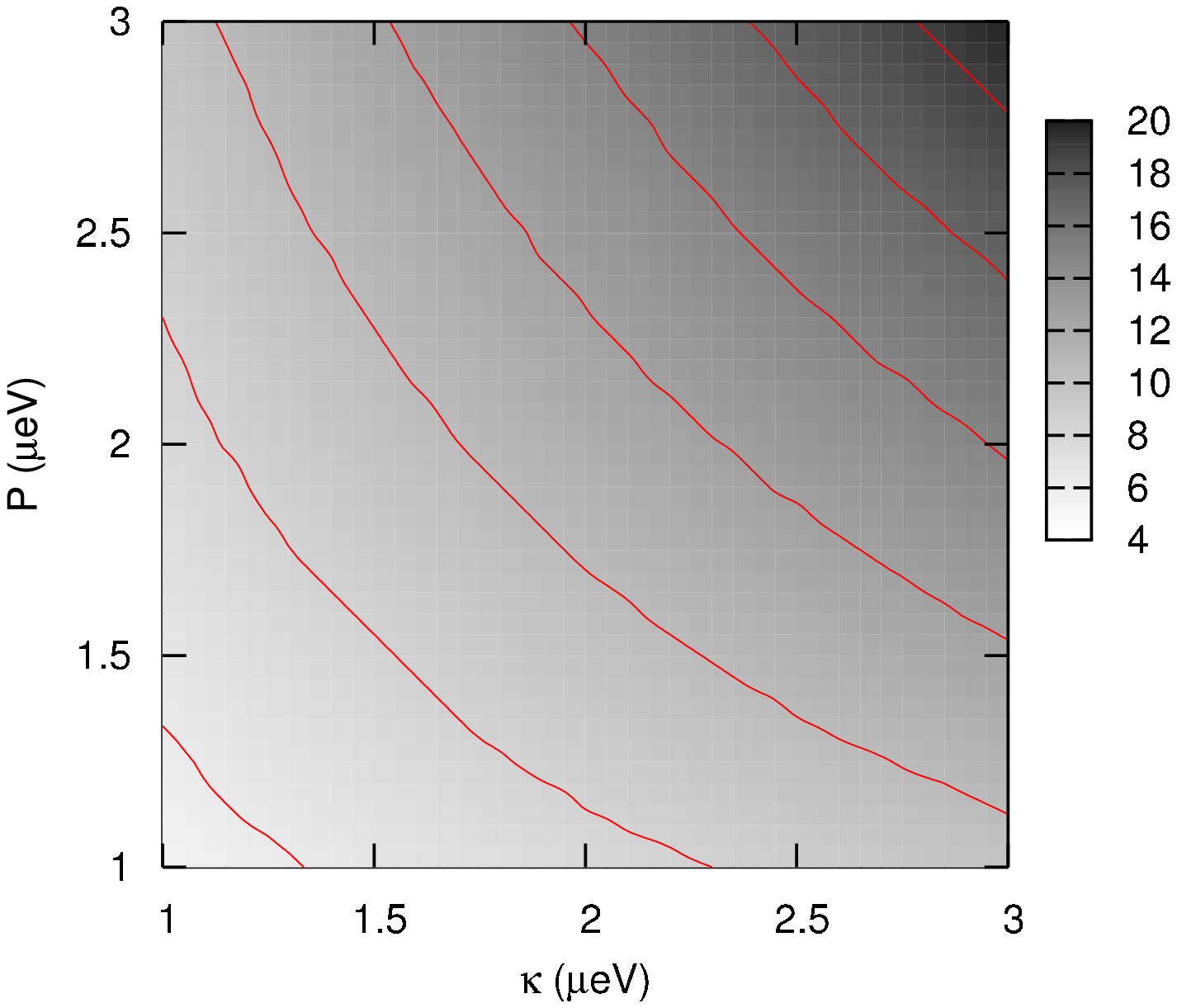}
  \end{minipage}
  \caption{\label{fig:8} Revivals time $\delta t_1$ (left panel),
  $\delta t_2$ (right panel) in ps as functions of $\kappa$ and
  $P$. As dissipative factors increase the elapsed time in the
  concurrence revivals becomes longer.}
 \end{figure}

\begin{figure}[!ht]
  \begin{minipage}{0.3\textwidth}
    \centering $t_1$
  \end{minipage}
  \ \hfill 
  \begin{minipage}{0.3\textwidth}
    \centering $t_2$
  \end{minipage}
  \ \hfill
  \begin{minipage}{0.3\textwidth}
    \centering $t_3$
  \end{minipage}
  \\
  \begin{minipage}{0.3\textwidth}
    \includegraphics[width=0.8\textwidth,angle=-90]{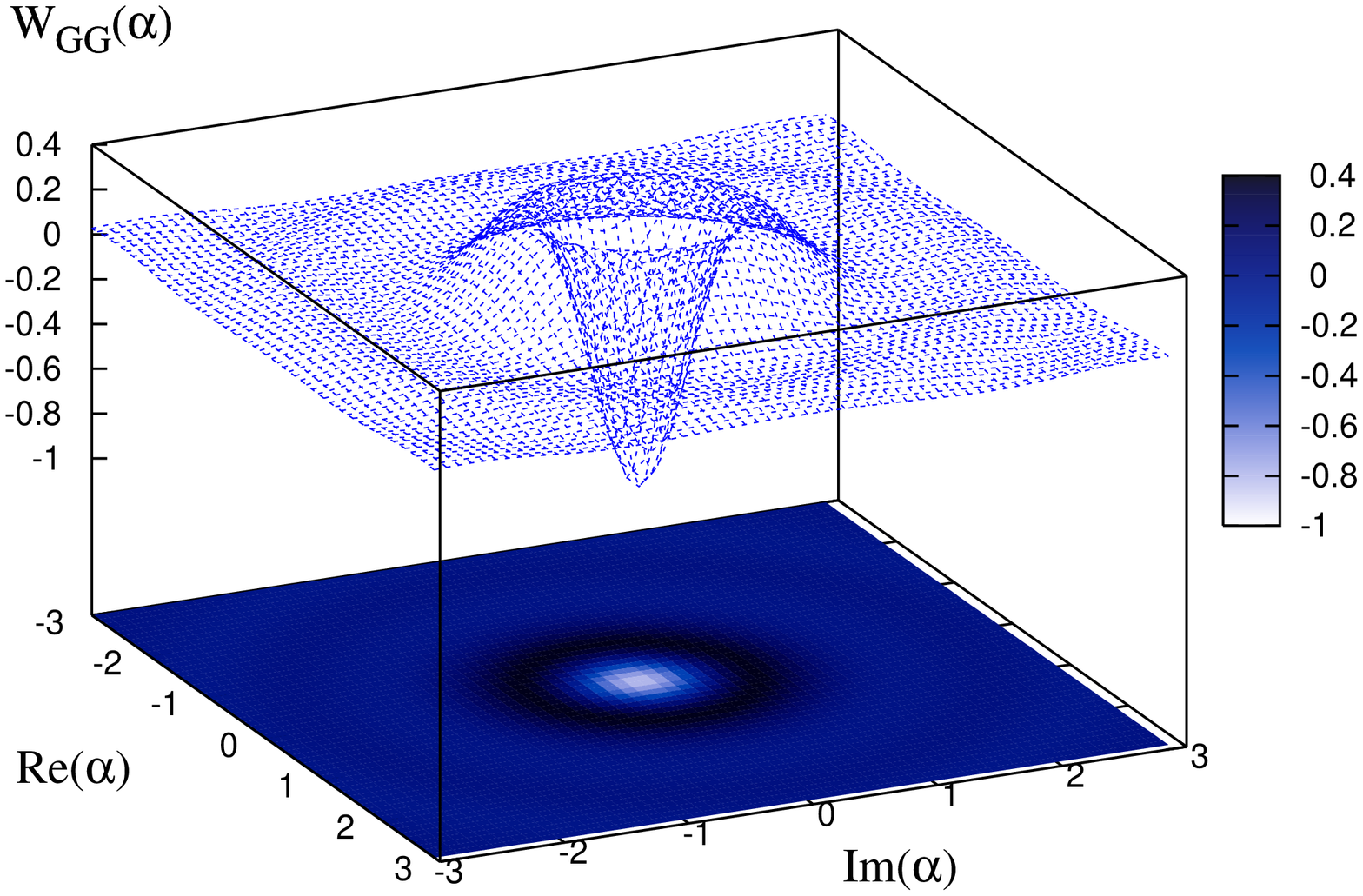}
  \end{minipage}
  \ \hfill 
  \begin{minipage}{0.3\textwidth}
    \includegraphics[width=0.8\textwidth,angle=-90]{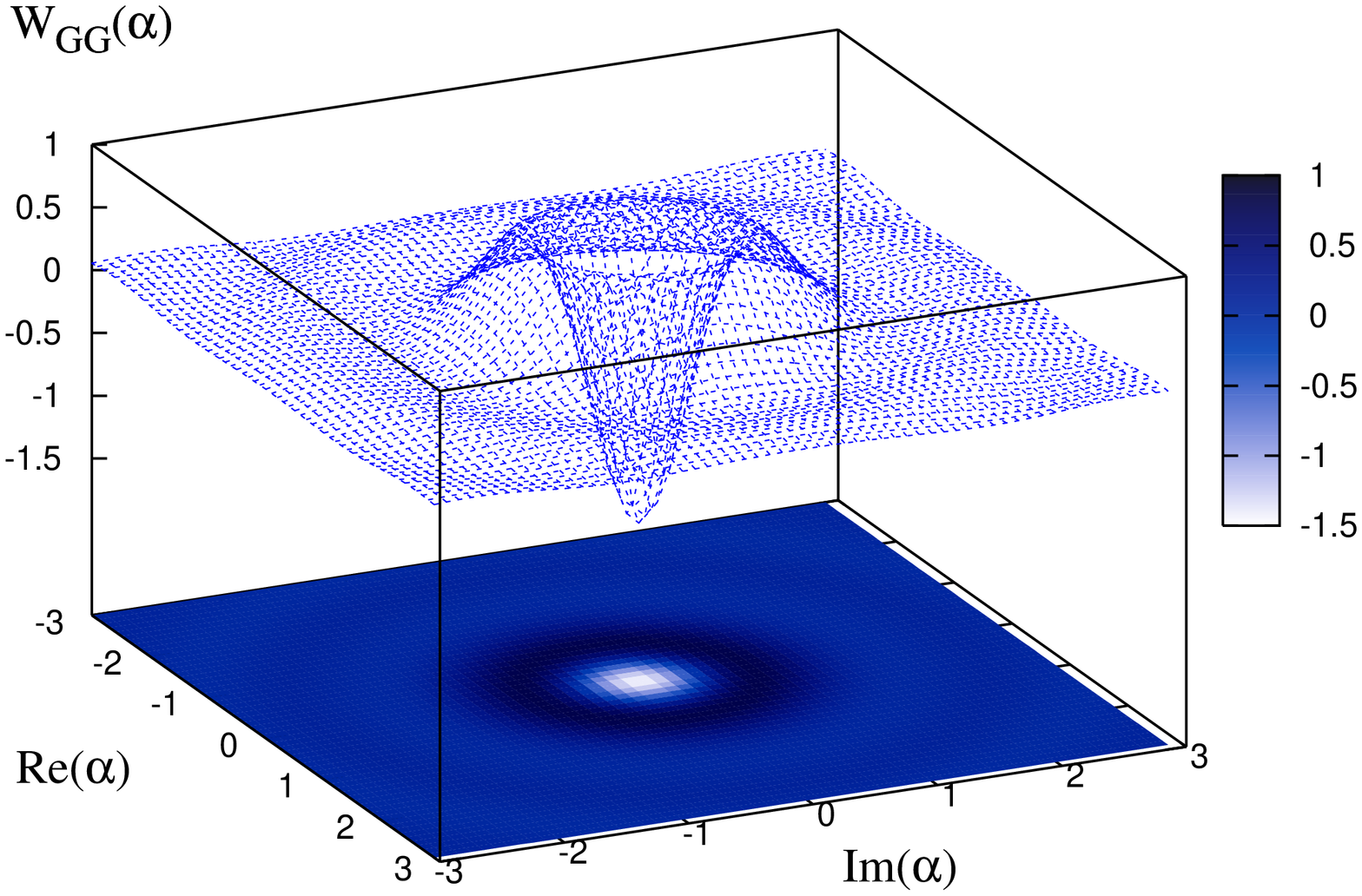}
  \end{minipage}
  \ \hfill
  \begin{minipage}{0.3\textwidth}
    \includegraphics[width=0.8\textwidth,angle=-90]{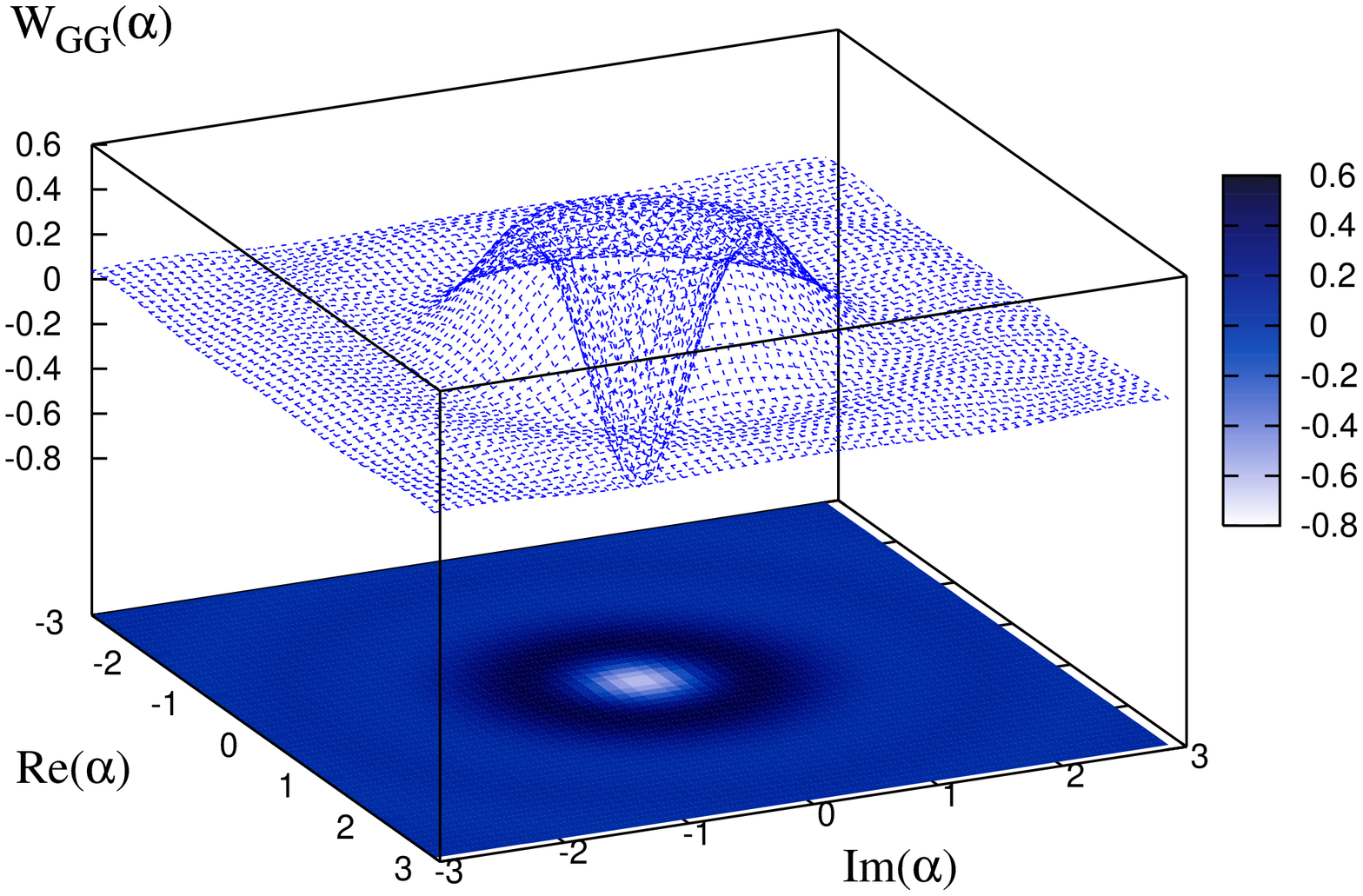}
  \end{minipage}
  \\
  \begin{minipage}{0.3\textwidth}
    \includegraphics[width=0.8\textwidth,angle=-90]{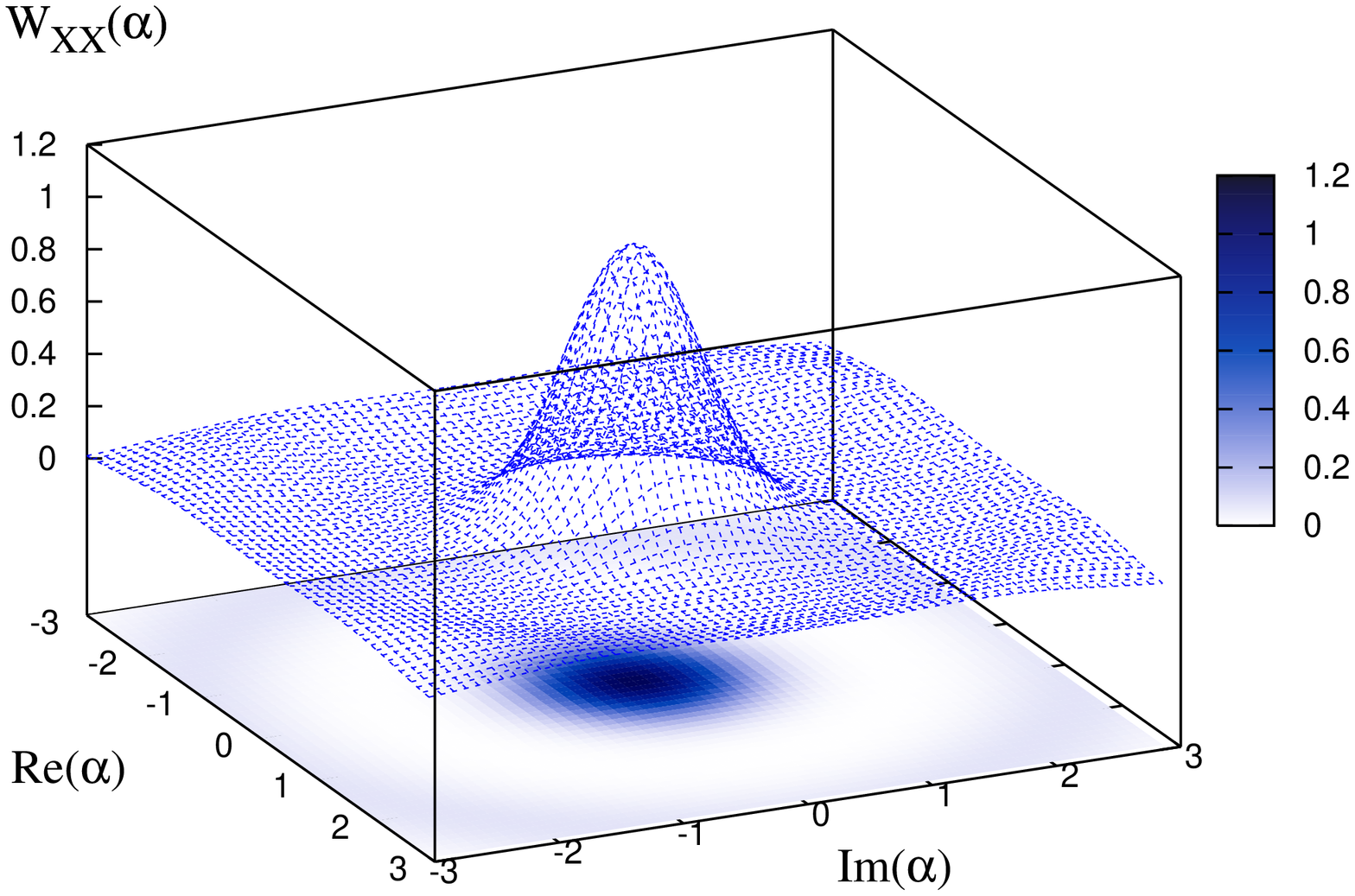}
  \end{minipage}
  \ \hfill 
  \begin{minipage}{0.3\textwidth}
    \includegraphics[width=0.8\textwidth,angle=-90]{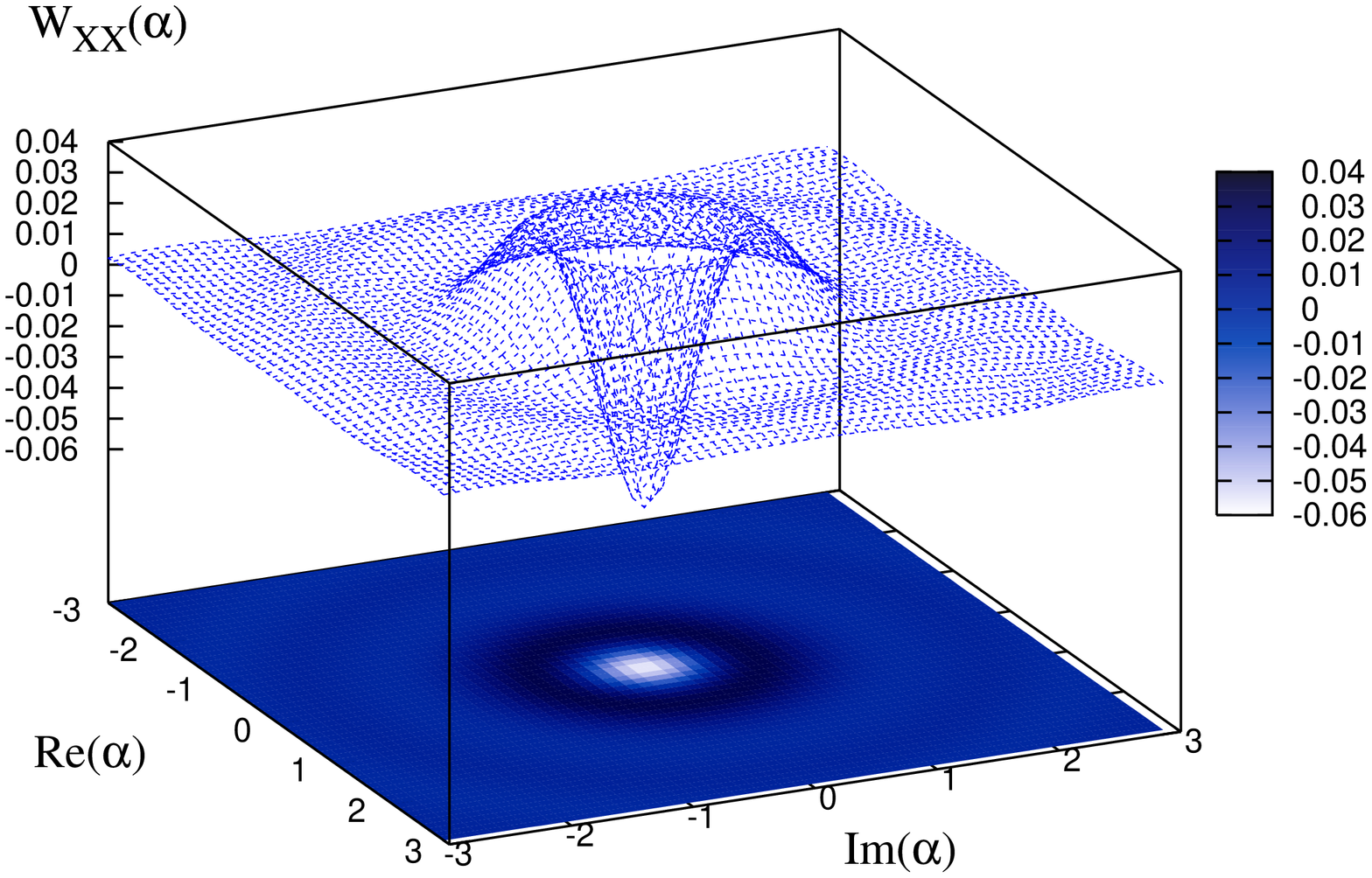}
  \end{minipage}
  \ \hfill
  \begin{minipage}{0.3\textwidth}
    \includegraphics[width=0.8\textwidth,angle=-90]{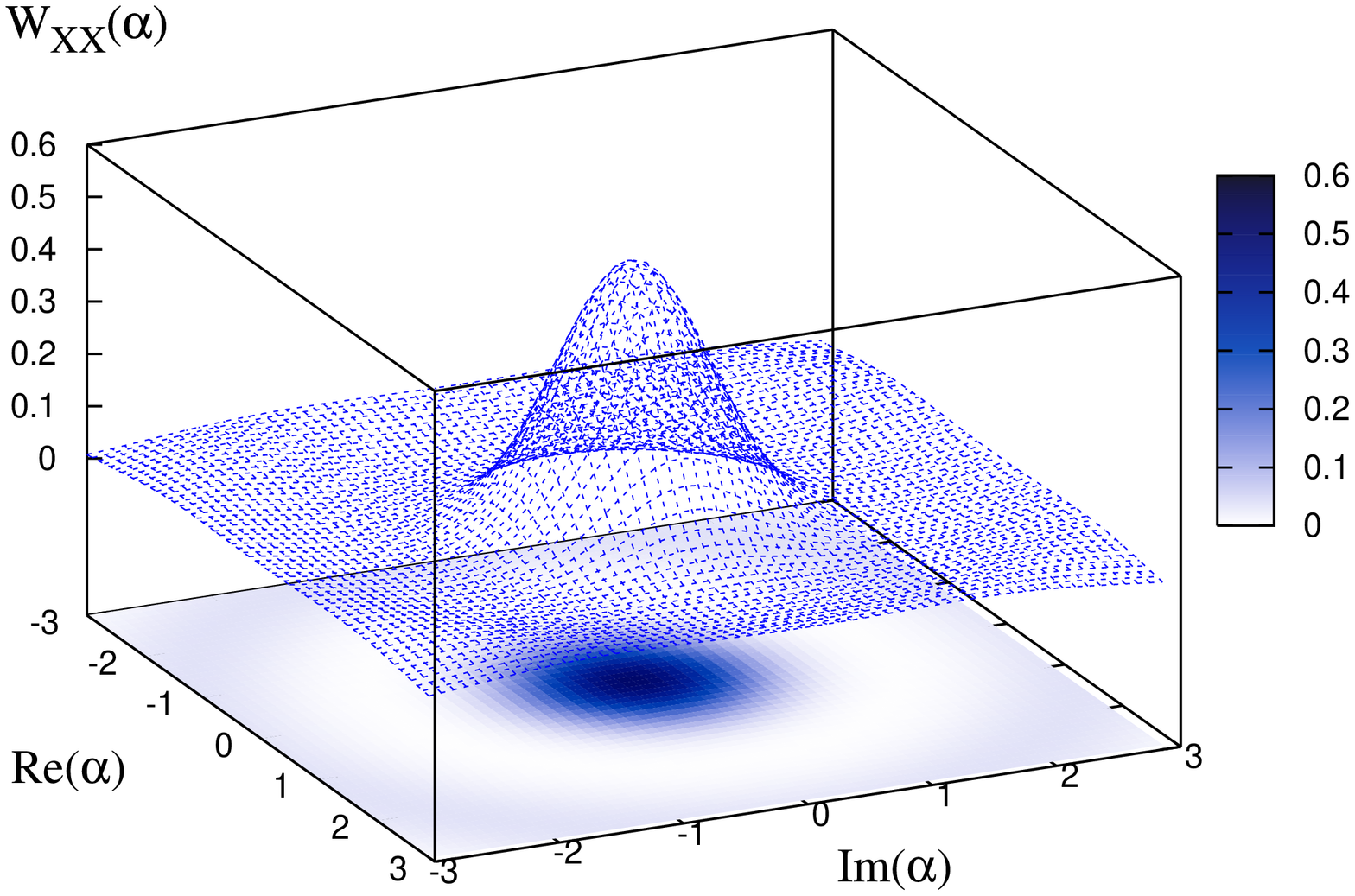}
  \end{minipage}
  \\
  \begin{minipage}{0.3\textwidth}
    \includegraphics[width=0.8\textwidth,angle=-90]{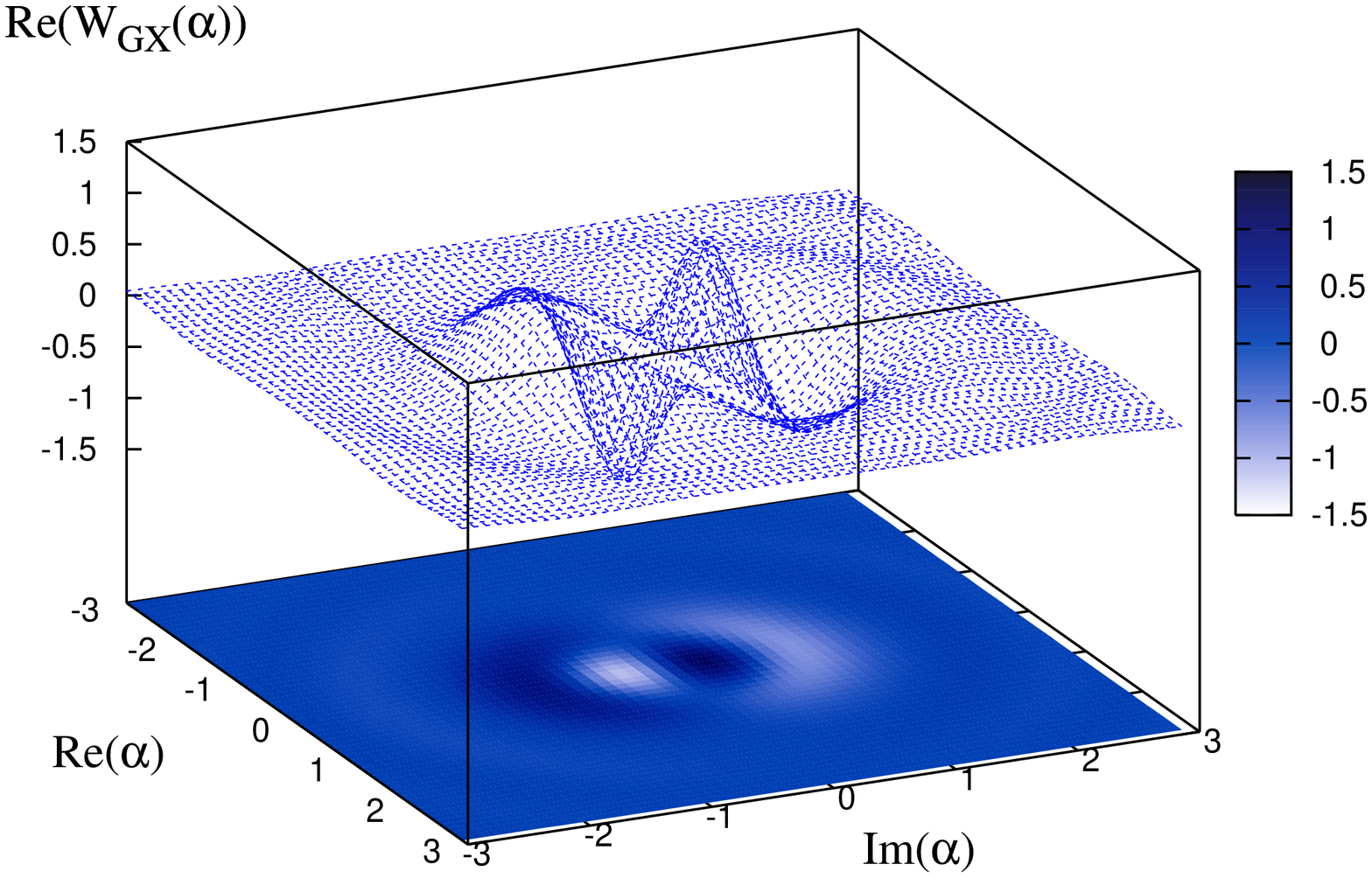}
  \end{minipage}
  \ \hfill 
  \begin{minipage}{0.3\textwidth}
    \includegraphics[width=0.8\textwidth,angle=-90]{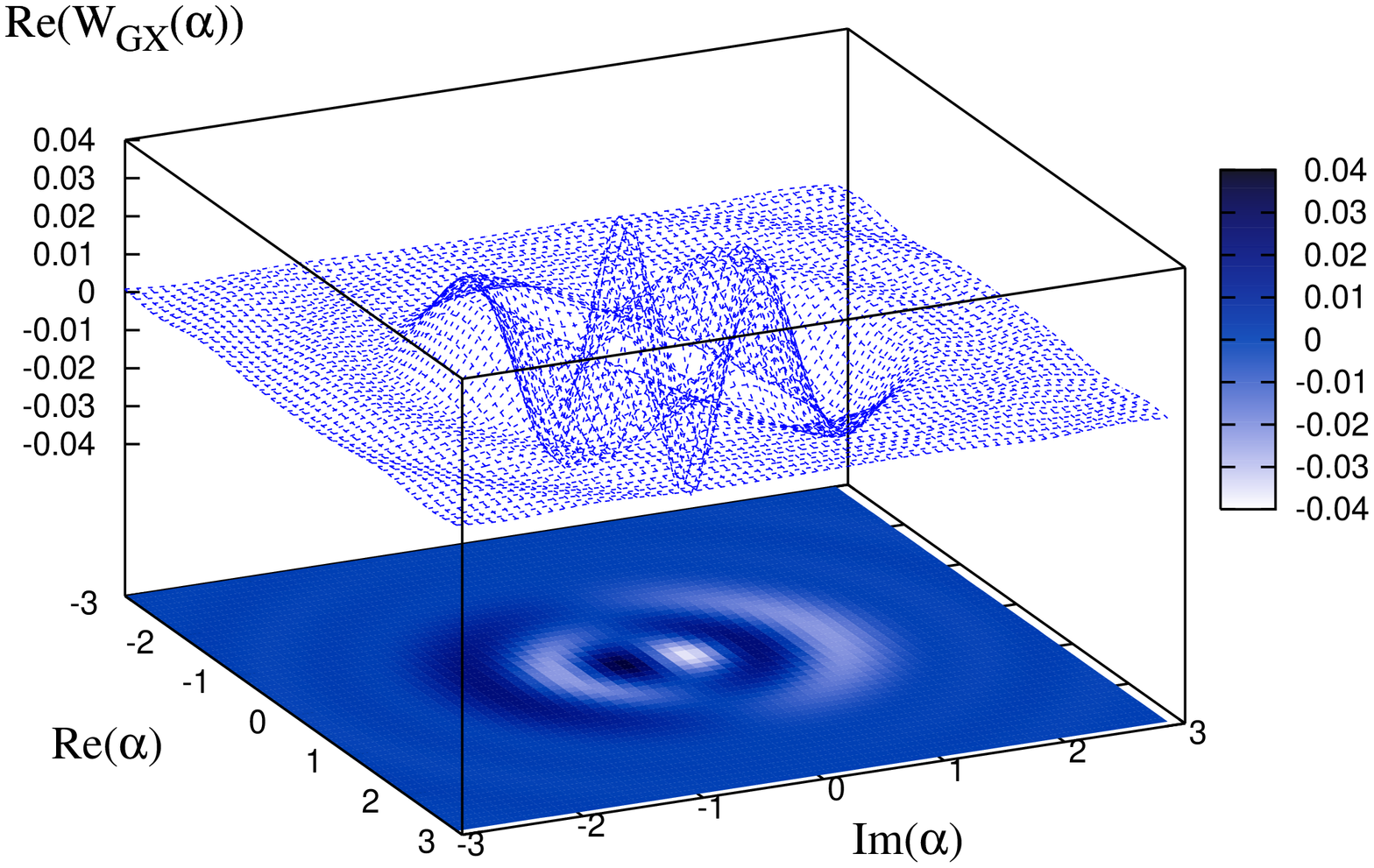}
  \end{minipage}
  \ \hfill
  \begin{minipage}{0.3\textwidth}
    \includegraphics[width=0.8\textwidth,angle=-90]{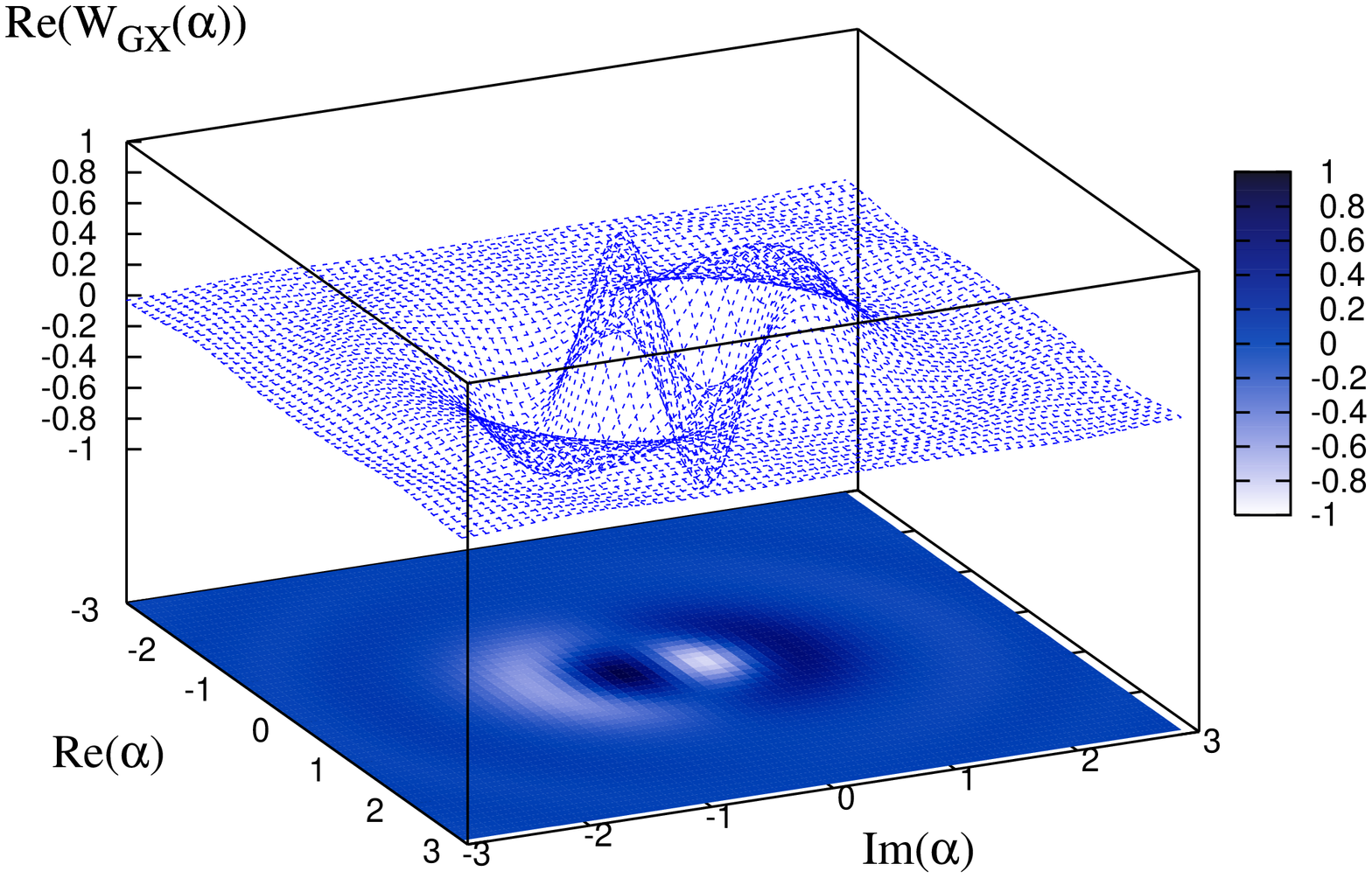}
  \end{minipage}
  \\
  \begin{minipage}{0.3\textwidth}
    \includegraphics[width=0.8\textwidth,angle=-90]{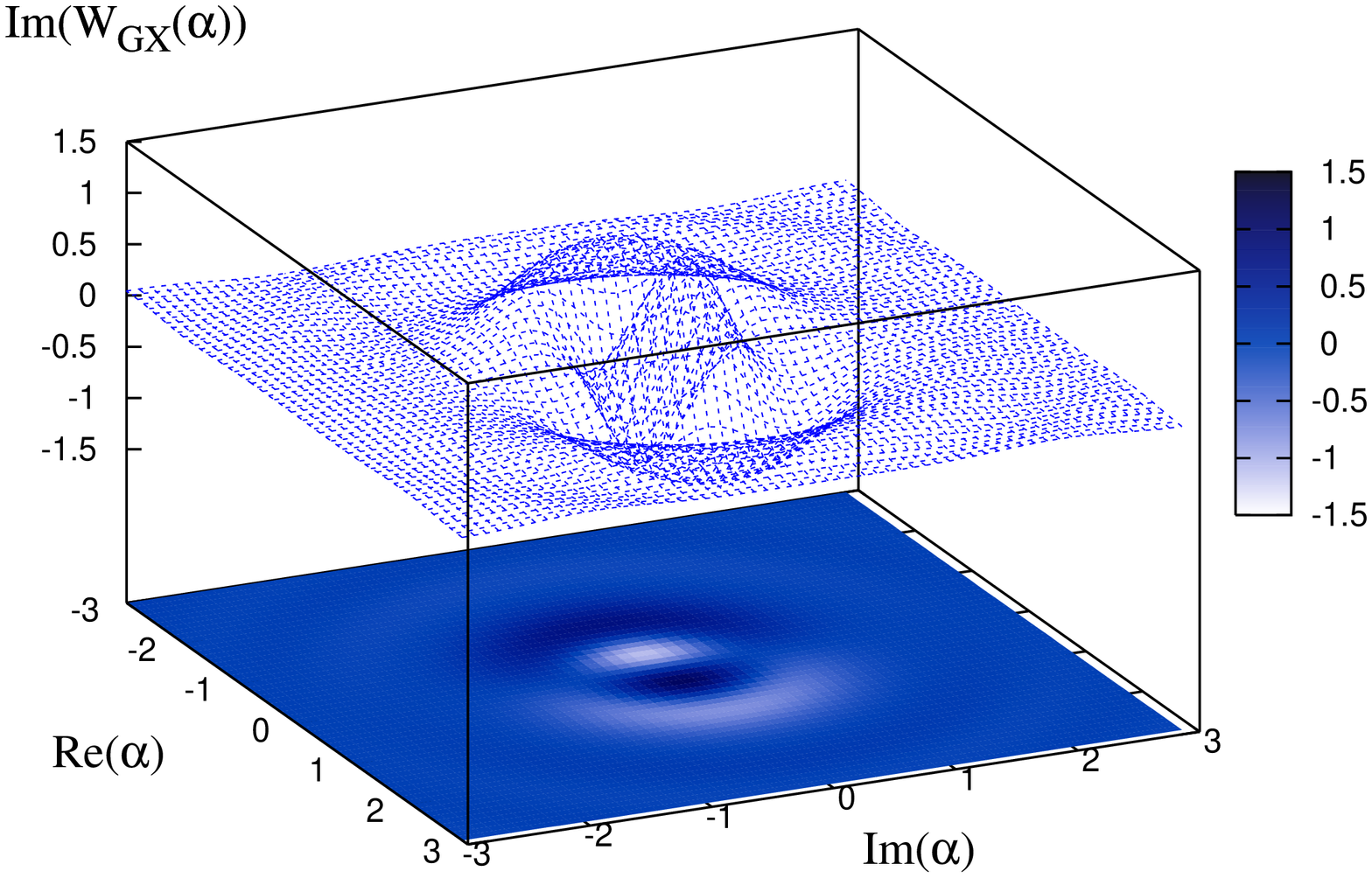}
  \end{minipage}
  \ \hfill 
  \begin{minipage}{0.3\textwidth}
    \includegraphics[width=0.8\textwidth,angle=-90]{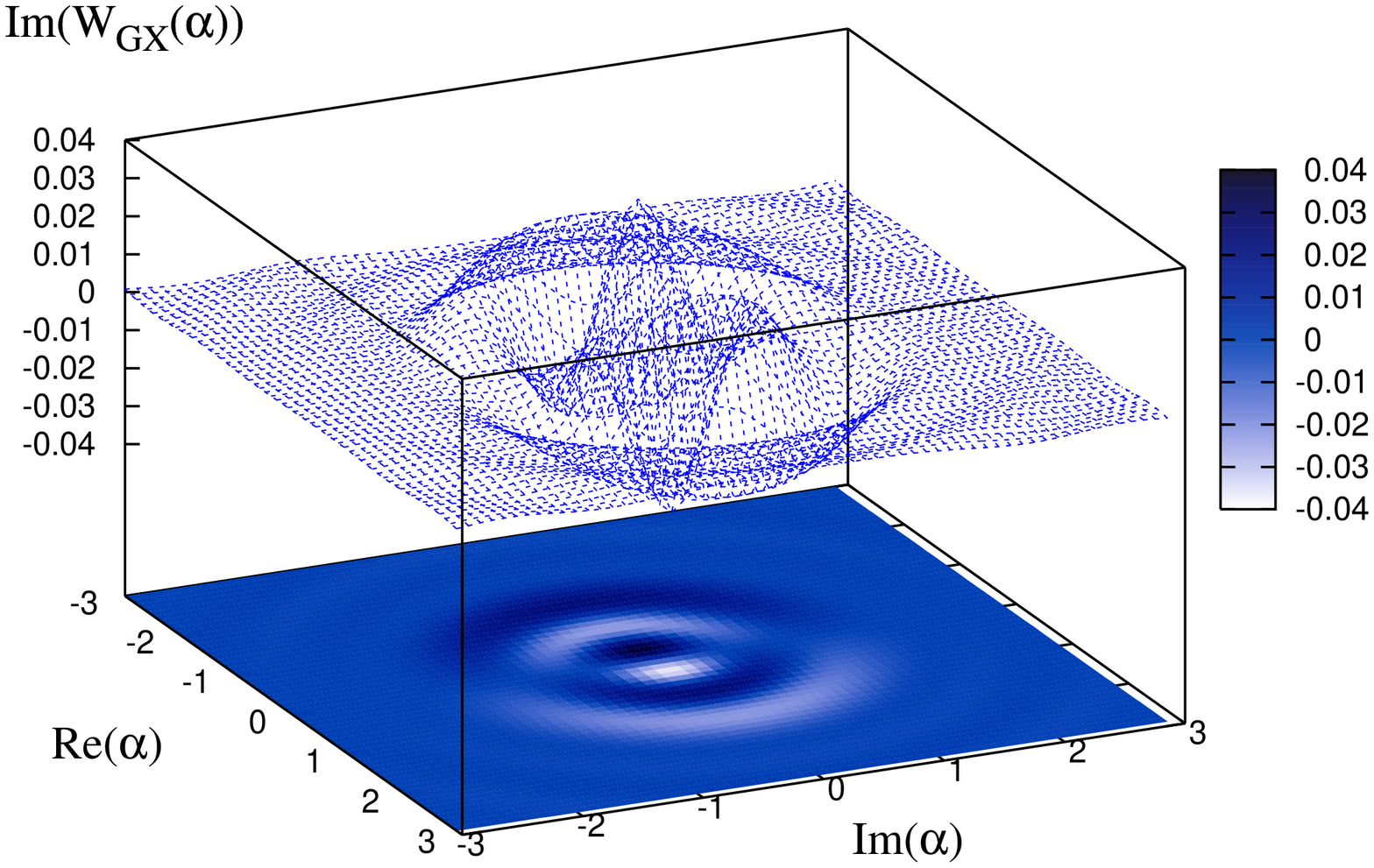}
  \end{minipage}
  \ \hfill
  \begin{minipage}{0.3\textwidth}
    \includegraphics[width=0.8\textwidth,angle=-90]{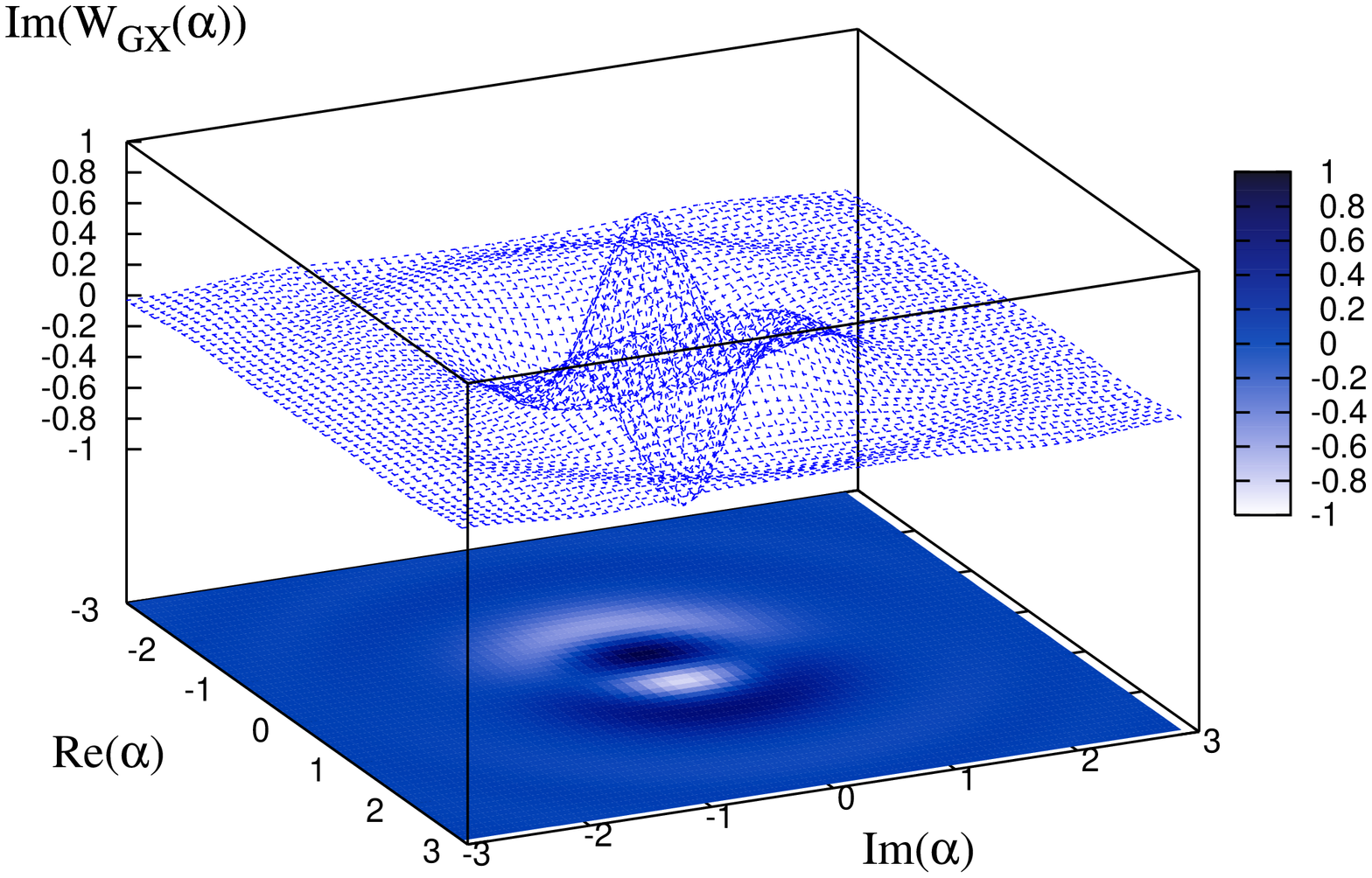}
  \end{minipage}
\caption{\label{fig:9} Time behavior for Wigner function matrix
elements at different concurrence regimes. Notice that in the times
of non vanishing concurrence  ($t_1$ and $t_3$) the shape of the matrix
elements is quite different, in particular the non diagonal elements
can not be obtained by multiplying the trace (which is a real valued
function) since they have more than one lobule. On the other hand in
the vanishing concurrence case ($t_2$) the shape is the same. \\
To further quantify the separability of the quantum state of the whole
system we can define the following quantity: $\delta W_{ij} (\alpha) = W_{ij}(\alpha) -
\frac{W_{ij}(0)}{W(0)}W(\alpha)$ if the state is separable then $\delta W_{ij} (\alpha)=0$ $\forall
\alpha$. Actually when one calculates $\delta W_{ij} (\alpha)$ in $t_1$ and $t_3$ $\delta
W_{ij} (\alpha) $ is bounded as follows $0 \leq |\delta
W_{ij} (\alpha)| \leq 3$ and in $t_2$ is has the following bound $0 \leq |\delta W_{ij}
(\alpha)| \leq 0.04$.
The times $t_1$, $t_2$ and $t_3$ are indicated in figure \ref{fig:7}.
}
\end{figure}

In the previous section a study of the relation between
dynamics and photoluminescence   was made, leading to a direct
identification of strong and weak coupling regimes. Now we want to study the behavior of the entanglement in these regimes. In order to do so
we now turn our attention to the simplified system described in section
\ref{sec02:sub04}. First we study the system in strong coupling regime with a
set of parameters as those described in \ref{sec03:WC}; however, since we are
mainly interested in the effects that dissipation has on the
entanglement we set $\Delta = 0$. Since spontaneous emission processes
are small as compared with the other effects, we also set $\gamma =
0$. Taking $ g=25\;\mu$eV we obtain the results in the upper panels of
fig.~\ref{fig:7}. In the middle left panel the evolution of the 
concurrence as a function of detuning is shown. Notice that when the
system moves from strong coupling to weak coupling 
%(dispersive regime\cite{Nemes})
the entanglement becomes larger, leading to a
non vanishing value of the concurrence for all times.\\

The range of parameters for this situation was carefully
chosen since great variations of them lead to unphysical results. Indeed,
if $\kappa$ is increased the cavity eventually will get empty, if $P$
is increased the basis is not large enough to describe the system,
since the mean number of photons becomes larger than 1.\\

Results on the evolution of concurrence show that it does not distinguish between
the two dissipative processes (parameters) considered ($\kappa$ and $P$). 
The reason for this behavior is that the concurrence of our system depends essentially of the
difference between the absolute value of the coherence term $\rho_{G1X0}$ and the square root
of the product of the populations $\rho_{G0G0}$ and $\rho_{X1X1}$. Notice for example that in figure \ref{fig:7} panel (e)
the concurrence vanishes for precisely the times when $|\rho_{G1X0}|<\sqrt{\rho_{G0G0}\rho_{X1X1}}$
and that the difference between the concurrence and $|\rho_{G1X0}|$ in their minima grows as the coherence $\sqrt{\rho_{G0G0}\rho_{X1X1}}$ grows as seen in panel (f).
In the dynamics of the density matrix we are considering  it is seen that the coherences decay exponentially with rates proportional to $\kappa+P$ and that on the other hand $P$ tends to increase the population of $\rho_{X1X1}$ and $\kappa$ tends to increase the population of $\rho_{G0G0}$ so that the difference $|\rho_{G1X0}|-\sqrt{\rho_{G0G0}\rho_{X1X1}}$ does not distinguish the two processes.\\
Even more interesting is
the fact that as the dissipation increases the zones where concurrence
vanishes (the so called Entanglement Sudden Death \cite{david}) become wider, revivals in the concurrence become
more separated as dissipation increases. Note that
when dissipation increases the maxima of the concurrence are less
defined and eventually disappear. In order to quantify these two
effects we compute the temporal length of the first two collapses of the
concurrence ($\delta t_1$ is the time interval of the first collapse and $\delta t_2$ the
time interval of the second one). The results obtained are plotted in fig.~\ref{fig:8}. It is clearly seen
that as the non hamiltonian effects become important the length of both intervals gets longer  
and that for any $\kappa$ and $P$ the second interval $\delta t_2$ is wider than the first one $\delta t_1$.
This behavior can be understood as follows: as explained in the last paragraph the degree of entanglement between the subsystems is the difference $|\rho_{G1X0}|-\sqrt{\rho_{G0G0}\rho_{X1X1}}$ and the dynamical behavior of the coherence term is an oscillatory function times an exponential decaying function with decay rate proportional to the sum of $\kappa$ and $P$. In order to have entanglement the absolute value of $|\rho_{G1X0}|$ must be greater than $\sqrt{\rho_{G0G0}\rho_{X1X1}}$. Now notice that as $P$ or $\kappa$ are increased the oscillations gets more damped. Because of the damping the amplitude of the oscillations is smaller and the absolute value of the coherence has to be in a time $t$ closer to a maxima in order to be greater than $\sqrt{\rho_{G0G0}\rho_{X1X1}}$, that is, the finite time where there is no entanglement approaches to the time interval between to successive maxima of $|\rho_{G1X0}|$ as $\kappa$ or $P$ is increased. To understand the fact that the first interval of sudden death $\delta t_1$ is smaller than the second one $\delta t_2$, simply notice that for the times in the second interval the amplitude of $|\rho_{G1X0}|$ will be smaller than in the first one, so that the times when the entanglement will be different from zero will be closer to the maxima of $|\rho_{G1X0}|$ resulting in a wider interval.

This behavior suggest that by manipulating experimentally accessible parameters such as the pumping rate and
the quality factor of the cavity one can in principle make coherence control in the system.
Finally, it is worth to mention that due to the coupling with external reservoirs the system evolves to a non pure state as evidenced in figure (\ref{fig:7}) where we plot the linear entropy.

Now we compute the elements of the Wigner function $W_{GG}$ and $W_{XX}$
(see fig.~\ref{fig:9}) in representative zones of the concurrence
function for typical values of the parameters: $ \kappa = 3$ $\mu$eV and
$ P=2$ $\mu$eV, at the times $t_1$, $t_2$ and $t_3$ indicated in figure \ref{fig:7}. 
Note that in the regions where concurrence goes to zero, phase space is similar
(up to a multiplicative constant) as suggested by $W_{XX}$, $W_{GG}$,
Re$(W_{GX})$ and Im$(W_{GX})$, so that the photonic states are similar for
both ground and excited excitonic states and hence the photonic states
are separable from the excitonic part leading to a vanishing
entanglement. On the other hand for non vanishing concurrence it is
clear that photonic states can not be separated.\\

\section{\label{conc} Conclusions}
We have built a phenomenological model that is able to describe both weak and strong coupling an that accounts for \emph{temperature} effects
in a microcavity quantum dot system.
We have used concurrence criterion and the Wigner function to carry out the
entanglement analysis related with the dynamical regimes in a simple
dissipative model. The strong coupling regime shows a periodical
disentanglement that depends on the dissipation rates. On the other
hand, the weak coupling regime shows no complete dynamical losing of
entanglement. This relation between dynamical regimes and entanglement
sudden death has been investigated and we have shown that on the one hand both the 
coherent emission ($\kappa$) and the incoherent pumping ($P$) affect the time windows where
there is no entanglement. In the same way, as time goes by this windows become wider due to 
the damping in the dynamics of the non diagonal terms of the density matrix caused by $\kappa$ and $P$. \\
Finally, we would like to point out that our results suggest that one
can control the entanglement between the subsystems by manipulating external parameters
such as the pumping rate $P$ and the cavity quality factor (which is related to the emission rate $\kappa$).

\section*{\label{aknonw} Acknowledgments}
Authors acknowledge the experimental group of Universidade Federal de Minas Gerais, specially P.S.S. Guimaraes for his contribution
during the elaboration of this work. This work was supported by
Universidad de Antioquia and COLCIENCIAS.

\end{document}